\documentclass[conference]{IEEEtran}
\IEEEoverridecommandlockouts

\newcommand\Tstrut{\rule{0pt}{2.4ex}}         
\newcommand\Bstrut{\rule[-0.9ex]{0pt}{0pt}}   

\usepackage{textcmds}
\usepackage{bm}
\usepackage{amsmath,amssymb}
\usepackage{enumerate}
\usepackage{graphicx}
\usepackage{tikz}
\usetikzlibrary{shapes.geometric, arrows}
\usepackage{hhline}
\usepackage{longtable}	
\usepackage{nicefrac}
\usepackage{multicol}
\usepackage{float}
\usepackage{multirow}
\usepackage{epsf,psfrag,amssymb,amsfonts,color,cite,fancybox}
\usepackage{array}
\usepackage[mathscr]{eucal}
\usepackage{algorithmic}

\usepackage{textcomp}
\usepackage{diagbox}
\usepackage{xcolor}
\usepackage{soul}
\usepackage{enumitem}
\allowdisplaybreaks

\usepackage{optidef}

\ifCLASSOPTIONcompsoc
\usepackage[caption=false,font=normalsize,labelfon
t=sf,textfont=sf]{subfig}
\else
\usepackage[caption=false,font=footnotesize]{subfi
g}
\fi

\DeclareMathOperator*{\E}{\mathbb{E}}

\DeclareMathOperator{\Std}{Std}

\IEEEoverridecommandlockouts

\makeatletter
  \def\my@tag@font{\normalsize}
  \def\maketag@@@#1{\hbox{\m@th\normalfont\my@tag@font#1}}
  \let\amsmath@eqref\eqref
  \renewcommand\eqref[1]{{\let\my@tag@font\relax\amsmath@eqref{#1}}}
\makeatother

\tikzstyle{startstop} = [rectangle, rounded corners, minimum width=1.2cm, minimum height=0.8cm,text centered, draw=black, fill=white!20]
\tikzstyle{io} = [trapezium, trapezium left angle=70, trapezium right angle=110, minimum width=0cm, minimum height=0cm, text centered, draw=black, fill=red!30]
\tikzstyle{process} = [rectangle, minimum width=0cm, minimum height=0cm, text centered, draw=black, fill=orange!15, rounded corners]
\tikzstyle{decision} = [diamond, minimum width=1.2cm, minimum height=0.8cm, text centered, draw=black, fill=blue!10]
\tikzstyle{arrow} = [thick,->,>=stealth]

\pdfoutput=1
\begin{document}
%
\title{A Data-Driven Energy Storage System-Based Algorithm for Monitoring the Small-Signal Stability of Power Grids with Volatile Wind Power \\
\thanks{This work is supported by the Fonds de Recherche du Qu\'{e}bec - Nature et technologies under Grant FRQ-NT PR-253686 and the Natural Sciences and Engineering Research Council (NSERC) under Discovery Grant NSERC RGPIN-2016-04570.}}


%
%

 \author{
 \IEEEauthorblockN{Ilias Zenelis, Georgia Pierrou, and Xiaozhe Wang}
 \IEEEauthorblockA{Department of Electrical and Computer Engineering, McGill University, Montreal, QC H3A 2K6, Canada\\
 ilias.zenelis@mail.mcgill.ca, georgia.pierrou@mail.mcgill.ca,  xiaozhe.wang2@mcgill.ca}
 }

\maketitle

\begin{abstract}

In this paper, we propose a data-driven energy storage system (ESS)-based method to enhance the online small-signal stability monitoring of power networks with high penetration of intermittent wind power. To accurately estimate inter-area modes that are closely related to the system's inherent stability characteristics, a novel algorithm that leverages on recent advances in wide-area measurement systems (WAMSs) and ESS technologies is developed. It is shown that the proposed approach can smooth the wind power fluctuations in near real-time using a small additional ESS capacity and thus significantly enhance 
the monitoring of small-signal stability. Dynamic Monte Carlo simulations on the IEEE 68-bus system are used to illustrate the effectiveness of the proposed algorithm in smoothing wind power
and estimating the inter-area mode statistical properties.


\end{abstract}


\section{Introduction}\label{1}
Wind power is a rapidly evolving renewable energy technology worldwide because of its cleanness, abundance and cost-effectiveness.
However, the volatile and stochastic nature of wind
poses many challenges to the secure operation of modern grids.
Concerning small-signal stability, the intermittency of wind energy may not only lead to instability \cite{Bu15} but also result in great obstacles regarding the power system stability monitoring and assessment \cite{Bu12}. Particularly,
poorly-damped inter-area oscillations
may become undetectable if small-signal stability monitoring is deteriorated. Such a phenomenon can cause major power outages \cite{Andersson05}. Recent works have proposed probabilistic approaches to study and quantify the impact of wind power on small-signal stability analysis, which may nevertheless require large computational effort \cite{Huang13}, \cite{Bazargan10}.

The recent advances in energy storage systems (ESSs) have
provided power engineers with an effective means to minimize the unwanted impacts of wind energy on power networks by smoothing wind power variations \cite{Sumper12}. However, even with ESSs, additional challenges may arise as power grids are transforming into large-scale networks with increasing complexity, due to the continuous integration of power electronics-based devices, the transmission system expansion, etc. In fact, \cite{Sheng20} has shown that the conventional model-based methods for small-signal stability monitoring may fail when the power system experiences unexpected disturbances or undetected topology changes. In this direction, measurement-based methods have been proposed in the recent literature to monitor and control small-signal stability
 considering load uncertainties \cite{Sheng20,Zenelis18,Zenelis20,Weng20}. 
 These strategies mainly rely on the enormous growth of wide-area measurement
systems (WAMSs) and phasor measurement units (PMUs) over the last $10-20$ years \cite{Kamwa01}. 
Despite of providing advancements, 
these works do not consider wind stochasticity.

To address these challenges, in this paper, a novel data-driven ESS-based algorithm for monitoring the small-signal stability of power grids with volatile wind power is proposed. Our method exploits two of the key emerging technologies--WAMS and ESS--that have already been massively installed in most power networks \cite{PNNL}, to accurately monitor the inter-area modes of systems with random renewables in near real-time. Particularly, we install ESSs at the wind generator side to smooth out the random wind power fluctuations. The assumption of having ESSs at wind generator side may be optimistic now, but is commonly made in the literature due to the fast rate of ESS deployment in power grids \cite{Sumper12,PNNL, Hasanien14}. 
We then apply an online data-driven mode identification approach to estimate the dynamic system state matrix and the inter-area
mode characteristics. Unlike probabilistic stability assessment approaches, the proposed method enables small-signal stability assessment online (within a $5$ minutes window). 
In addition, it will be 
shown that the ESS capacity required for wind power smoothing can be determined based on the statistical properties of wind farm power output, whereas its size is not significant when compared with the large scale of power grids. 
To the best of
authors'
knowledge, this work represents the first attempt to enhance  the
data-driven
monitoring
of small-signal
stability considering the stochastic
nature of renewable energy sources.
\vspace{-2pt}
\section{Stochastic Model for Power Grid Dynamics}\label{2}
Inter-area modes lie in the low-frequency portion of the electromechanical mode spectrum (i.e. $0.1 - 1$ Hz) \cite{Kundur94}. Thus, fast generator dynamics can be neglected and aggregated synchronous machines can be represented by the classical model \cite{Kundur94}.
By numbering generator buses as $i=1,...,n$:
\begin{eqnarray}
\dot{\delta}_{i} &=& \omega_{i} \\
M_{i}\dot{\omega}_{i} &=& P_{m_{i}}- P_{e_{i}} - D_{i}\omega_{i} \label{swing2}\\
P_{e_{i}} &=& E_{i}\sum\limits_{j=1}^n E_{j}|Y_{i,j}|\cos(\delta_{i}-\delta_{j}-\phi_{i,j}) \label{swing3}
\end{eqnarray}
where $\delta_{i}$ is the rotor angle, $\omega_{i}$ the rotor speed deviation from synchronous speed, $M_{i}$ the inertia coefficient, $D_{i}$ the damping coefficient, $P_{m_i}$ the mechanical power input, $P_{e_i}$ the electrical power output, $E_i$ the transient emf magnitude, and $|Y_{i,j}|\angle{\phi_{i,j}}$ the
$\textit{(i,j)}^{th}$ entry of the Kron-reduced admittance
matrix $Y$.

\subsection{Stochastic Load Model}

Generator dynamics prevail over load dynamics in the study of inter-area modes. Therefore, we model loads as constant impedances to simplify the computations and obtain the generator electromechanical dynamics from the network dynamics
\cite{Kundur94}. Considering a steady-state grid operation, we assume that inter-area modes are excited by Gaussian load fluctuations
that translate into variations of the diagonal elements of $Y$ \cite{Theresa13}:
\vspace{-4pt}
\begin{equation}
\vspace{-4pt}
\label{load_variation}
{Y_{i,i}}' = |Y_{i,i}|(1 + \sigma_{i}\xi_i)\angle{\phi_{i,i}}, \quad  i=1,...,n
\end{equation}
where $\xi_{i}$
are mutually independent standard Gaussian random variables, $\sigma_{i}$ is the standard deviation of load variations and $Y_{i,i}\angle{\phi_{i,i}}=G_{i,i}+jB_{i,i}$ is the $\textit{(i,i)}^{th}$ element of $Y$. Substituting (\ref{load_variation}) into (\ref{swing3}), i.e. replacing $|Y_{i,i}|$ with $|Y_{i,i}|(1 + \sigma_{i}\xi_i)$
gives
\vspace{-2pt}
\begin{equation}\label{new_Pe}
\vspace{-2pt}
{P_{e_{i}}}'={P_{e_{i}}}+E_{i}^2G_{i,i} \sigma_{i}\xi_i
\end{equation}


\subsection{Stochastic Wind Speed Model}\label{II-C}

In this work, wind farms associated to a wind speed model are integrated into the power network.
Due to its intermittency, wind speed adds stochastic perturbations to the grid that can be described by various continuous probability distributions, such as the Weibull distribution, the beta distribution, etc \cite{Carta09}. Therefore, wind speed can be statistically represented by a generic vector stochastic process $\bm {v_w}=[v_{w_1},...,v_{w_m}]^T$ where $m$ is the number of wind farms installed in the grid. In the simulation study of this paper, wind speed is modeled as  a Weibull distributed stochastic process by a set of stochastic differential algebraic equations, following \cite{Milano13, Wang15, Pierrou19}. 

The power captured by a variable speed wind farm is
\begin{equation}
    P_{w_j} = \frac{n_{g_j}\rho}{2}c_{p_j}A_{r_j}{v_{w_j}}^3, \quad j=1,...,m \label{Pwind}
\end{equation}
where $n_{g_j}$ is the number of wind turbines that compose the wind farm, $\rho$ is the air density, $c_{p_j}$ is the performance coefficient,
and $A_{r_j}$ is the turbine rotor swept area. Hence, wind farm power output is also a stochastic process.
Wind power dynamics are closely coupled to the voltage phasors of the buses where wind farms are installed \cite{Slootweg03}, 
which subsequently affect the electrical power output $P_{e_{i}}$
of synchronous generators (see (\ref{swing3})). That being said,
$P_{e_{i}}$, $i=1,...,n$ is a function of $\bm{v_w}$. Thus, (\ref{new_Pe}) can be re-written as:
\begin{equation}\label{new_Pe_eta}
{P_{e_{i}}}'={P_{e_{i}}}(\bm{v_w})+E_{i}^2G_{i,i} \sigma_{i}\xi_i
\end{equation}

\subsection{Stochastic Dynamic Power System Model}

Substituting (\ref{new_Pe_eta}) to (\ref{swing2}) , we obtain the power system dynamic model operating around steady state under the influence of small random load fluctuations and wind speed perturbations:
\begin{eqnarray}
\dot{\delta}_{i} &=& \omega_{i} \label{swing1rand}\\
M_{i}\dot{\omega}_{i} &=& P_{m_{i}}- {P_{e_{i}}}(\bm{v_w}) - D_{i}\omega_{i}- E_{i}^{2}G_{i,i}\sigma_{i}\xi_{i} \label{swing2rand}
\end{eqnarray}
Linearization around the stationary point $(\bm{\delta_0}, \bm{\omega_0})$ yields:
\begin{equation}
\label{matrix_form}
\begin{gathered}
\resizebox{1\hsize}{!}{$
\underbrace{
\begin{bmatrix}
    \Delta\dot{\bm{\delta}}\\
    \Delta\dot{\bm{\omega}}
\end{bmatrix}}_{\dot{\bm x}}
=
\underbrace{
\begin{bmatrix}
  0_{n\times n} & I_{n\times n}\\
  -M^{-1}\frac{\partial {\bm {P_{e}}}(\bm{v_w})}{\partial {\bm \delta}} & -M^{-1}D
\end{bmatrix}}_{A}
\underbrace{
\begin{bmatrix}
    \Delta\bm{\delta}\\
    \Delta\bm{\omega}
\end{bmatrix}}_{{\bm x}}
+\underbrace{\begin{bmatrix}
0_{n\times n}\\
-M^{-1}E^2G\Sigma
\end{bmatrix}}_{B_\xi}
{\bm \xi}$}
\end{gathered}
\vspace{-5pt}
\end{equation}
where $\bm{\delta}=[\delta_{1},...,\delta_{n}]^T$, $\bm{\omega}=[\omega_{1},...,\omega_{n}]^T$, $\Delta\bm{\delta}=\bm{\delta}-\bm{\delta_0}$, $\Delta\bm{\omega}=\bm{\omega}-\bm{\omega_0}$,
$\bm{\xi} = [\xi_{1},...,\xi_{n}]^T$, $\bm{P_{e}}(\bm{v_w})=[P_{e_{1}}(\bm{v_w}),...,P_{e_{n}}(\bm{v_w})]^T$, $G=\mbox{diag}([G_{1,1},...,G_{n,n}])$, $M=\mbox{diag}([M_{1},...,M_{n}])$, $D=\mbox{diag}([D_1,...,D_n])$, $E=\mbox{diag}([E_{1},...,E_{n}])$, and $\Sigma=\mbox{diag}([\sigma_{1},...,\sigma_{n}])$.
 $B_\xi$
expresses
the effect of load variations. $A$ is the state matrix whose
eigenproperties
provide all the modal information including frequencies, damping ratios, etc.
Thus, the accurate knowledge of $A$ plays a crucial role in
identifying inter-area modes
and performing online small-signal stability monitoring. 
Traditionally, the calculation of $A$ requires the knowledge of  $\frac{\partial {\bm {P_{e}}}(\bm{v_w})}{\partial {\bm \delta}}$, $M$ and $D$. However, it may be hard to obtain the exact values of $M$ and $D$ in large-scale power grids \cite{Zenelis20}. Moreover, the computation of the Jacobian $\frac{\partial {\bm {P_{e}}}(\bm{v_w})}{\partial {\bm \delta}}$ requires information about the network model and its parameters (e.g. $Y$ and $E$),
which may be unknown or corrupted in practice \cite{Zenelis18}.

\subsection{Data-Driven Inter-Area Mode Estimation}
To overcome the aforementioned challenges, the purely data-driven strategy
\cite{Sheng20} can be exploited to estimate $A$, and thus the inter-area mode properties, from PMU data. For simplicity, we assume that all generator terminal buses are equipped with PMUs that provide measurements of real-time phasors of voltages and currents. However, the method can also handle cases of missing PMUs
\cite{Sheng20}. Rotor angles  $\bm{\delta}$ and speed deviations $\bm{\omega}$ can be estimated from synchrophasor data around steady state \cite{Zhou15}. Hence, the state vector ${\bm x}=[\Delta\bm{\delta},\Delta\bm{\omega}]^T$ is obtained. The stationary covariance matrix $C_{\bm x \bm x}$ and the $\tau$-lag time correlation matrix $G_{\bm x \bm x}\left( \tau  \right)$ of $\bm{x}$ satisfy:
\vspace{-3pt}
\begin{eqnarray}
C_{\bm x \bm x} &=& \E ({\left[ {{\bm{x}}\left( {t} \right) - \bar {\bm{x}} } \right], {{\left[ {{\bm{x}}\left( t \right) - \bar {\bm{x}} } \right]}^T}})\\
G_{\bm x \bm x}\left( \tau  \right) &=& \E( {\left[ {{\bm{x}}\left( {t + \tau } \right) - \bar {\bm{x}} } \right], {{\left[ {{\bm{x}}\left( t \right) - \bar {\bm{x}} } \right]}^T}})
\end{eqnarray}
According to the regression theorem for the Ornstein-Uhlenbeck process
$A$ can be estimated purely from the statistics of state variables that can be obtained from PMU data:
\vspace{-8pt}
\begin{equation}
    A = \frac{1}{\tau }\log \left[ G_{\bm x \bm x}\left( \tau  \right){C_{\bm x \bm x}}^{ - 1} \right]
    \label{eq:estA_c}
\end{equation}
which bypasses the knowledge of network topology and generator parameters.
After estimating $A$ purely from field measurements, we extract the inter-area mode information by modal analysis. Each inter-area mode is associated to an eigenpair $\lambda_{{i_\pm}} = r_i \pm jh_i$, $i\in \{1,...,n\}$ of $A$ and has its own frequency $f_i=\frac{h_i}{2\pi}$ and damping ratio $\zeta_i=\frac{-r_i}{\mid{\lambda_i}\mid}$.

When wind speed fluctuations are trivial (i.e. the variance of $\bm {v_w}$ is negligible), $P_{w_j}$ remains approximately constant and the normally distributed load randomness prevails over wind randomness. Since $A$ is almost fixed, 
$\bm{x}$ can be termed as a vector Ornstein-Uhlenbeck process 
in steady-state operation. Consequently, the data-driven method (\ref{eq:estA_c}) is expected to yield accurate inter-area mode estimation results that greatly enhance small-signal stability monitoring.
On the other hand, if wind speed variations become more significant, the non-Gaussian distributed wind randomness dominates load uncertainty. 
Meanwhile, $P_{w_j}$ might be highly fluctuating. As a result, $A$ is no longer constant such that the data-driven method \cite{Sheng20} based on the property of the vector Ornstein-Uhlenbeck process may fail to provide accurate estimation for inter-area modes. 
Nonetheless, the recent advancement of ESSs can be used to smooth out the random wind power, thus improving the inter-area mode estimation accuracy. 
It will be shown in the next section that by installing a small extra capacity to the existing ESS infrastructure, only a negligible additional effort is required to achieve an accurate real-time small-signal
stability monitoring of power grids with volatile wind power.




\subsection{Energy Storage System (ESS) for Wind Power Smoothing}

The randomness of wind power results in active power imbalance on the wind generator side:
\vspace{-2pt}
\begin{equation}\label{eq_Pimb}
    P_{im_j}(t_k)=P_{w_j}(t_k)-P_{ref_j}, \quad k=1,...,N
    \vspace{-2pt}
\end{equation}
where $P_{w_j}(t_k)$ is the actual power output, $P_{ref_j}$ is the reference (rated) power output, and $P_{im_j}(t_k)$ is the initial power imbalance of wind farm $j\in\{1,...,m\}$ at time instant $t_k$. 
To smooth the wind power fluctuations, we assume that every wind farm is equipped with an ESS \cite{Hasanien14}. 
Inspired by the single-bus multitimescale method introduced in \cite{Gamal13}, we model ESS discrete dynamics as follows:
\vspace{-2pt}
\begin{equation}\label{ESS_dyn}
    S_j(t_{k+1})=S_j(t_k)+\eta_{c_j}C_j(t_k)-\frac{1}{\eta_{d_j}}D_j(t_k)
    \vspace{-2pt}
\end{equation}
where
$S_j(t_k)\leq S_{max_j}$, $C_j(t_k)\leq C_{max_j}$, $D_j(t_k)\leq D_{max_j}$ are the stored, charging and discharging power of ESS $j\in\{1,...,m\}$ at time instant $t_k$, respectively; $\eta_{c_j}$ is the charging efficiency (ratio of charged to input power);
$\eta_{d_j}$ is the discharging efficiency (ratio of output to discharged power). $S_{max_j}$, $C_{max_j}$ and $D_{max_j}$ denote the ESS power capacity
for smoothing purposes, the maximum ESS charging power and the maximum ESS discharging power, respectively, 
while $S_j(t_1)$ is known. If $P_{im_j}(t_k)\geq0$, there is a surplus of energy at wind farm $j$ and the ESS is charged with $0 \leq C_j(t_k)\leq \min\{P_{im_j}(t_k),C_{max_j}\}$ and $D_j(t_k)=0$. 
If $P_{im_j}(t_k)\leq0$, there is a deficit of energy at wind farm $j$ and the ESS is discharged with $0 \leq D_j(t_k)\leq \min\{-P_{im_j}(t_k),D_{max_j}\}$ and $C_j(t_k)=0$.
The goal of ESS control (\ref{ESS_dyn}) is to minimize the
expected average magnitude of the
residual power imbalance
\begin{equation}
    P_{res_j}(t_k)=P_{im_j}(t_k)-C_j(t_k)+D_j(t_k)
\end{equation}
i.e. the wind power imbalance after ESS operation. In other words, $\mu_{P_{res_j}}=\E(\frac{1}{N}\sum_{k=1}^{N}|P_{res_j}(t_k)|)$ needs to be as close to zero as possible so as to give a smoothed wind farm power output. It can be proved that a greedy policy (i.e. charging/discharging sequence)
${\pi_o}^*=\{({C_j}^*(t_k),{D_j}^*(t_k)): k=1,...,N\}$
solves the aforementioned minimization problem optimally \cite{Gamal13}. Particularly, if $P_{im_j}(t_k)\geq0$ then:
\vspace{-3pt}
\begin{equation}
\label{policy1}
\footnotesize
{C_j}^*(t_k)=
    \begin{cases}
    C_{max_j}, \mbox{if } C_{max_j}\leq \min\{P_{im_j}(t_k), \frac{S_{max_j}-S_j(t_k)}{\eta_{c_j}}\}\\[5pt]
    P_{im_j}(t_k), \mbox{if } P_{im_j}(t_k)<\min\{ \frac{S_{max_j}-S_j(t_k)}{\eta_{c_j}},C_{max_j}\}\\[5pt]
    \frac{S_{max_j}-S_j(t_k)}{\eta_{c_j}}, \mbox{otherwise}
    \end{cases}
\end{equation}
\vspace{-5pt}
In contrast, if $P_{im_j}(t_k)\leq0$ then:
\begin{equation}\label{policy2}
\footnotesize
{D_j}^*(t_k)=
    \begin{cases}
    D_{max_j}, \mbox{if } \max\{P_{im_j}(t_k),-\eta_{d_j}S_j(t_k)\}<-D_{max_j}\\[5pt]
    -P_{im_j}(t_k), \mbox{if } \max\{-\eta_{d_j}S_j(t_k),-D_{max_j}\}\leq P_{im_j}(t_k)\\[5pt]
    \eta_{d_j}S_j(t_k), \mbox{otherwise}
    \end{cases}
\end{equation}

This ESS policy allows the accurate online small-signal stability monitoring \cite{Sheng20} by smoothing the wind farm power output variations. Note that the considered generic ESS could be pumped hydro storage, battery energy storage, etc \cite{Hug10}. Based on the above, we propose a data-driven ESS-based algorithm for monitoring the small-signal stability of power systems  with intermittent wind power 
(see Fig. \ref{chart}). Particularly:

\begin{figure}[!t]
\begin{center}
\begin{tikzpicture}[node distance=0.9cm]
[every label/.append style={text=red, font=\tiny}]
\tikzstyle{every node}=[font=\footnotesize]
\tikzstyle{every label}=[text=blue!60, font=\scriptsize]
\node (wind_speed) [process, label={[xshift=-2.05cm, yshift=0cm]\textbf{Step 1}}, xshift=2.3cm, fill=yellow!20] {Calculate $\Std({P_{im}}_j(t_k)), j=1,...,m$};
\node (check_stand) [decision, label={[xshift=-1.4cm, yshift=-0.6cm]\textbf{Step 2}}, below of =wind_speed, yshift=-0.8cm, aspect=2] {$\Std({P_{im}}_j(t_k))> \gamma_p$ {\large ?}};
\node (ESS_OFF) [io, right of =check_stand, xshift= 2.8cm] {ESS $j$ is OFF};
\node (ESS_ON) [io, below of =check_stand, yshift = -0.75cm, fill=green!20] {ESS $j$ is ON};
\node (find_Smax) [process, label={[xshift=-1.7cm, yshift=-0cm]\textbf{Step 3}}, below of=ESS_ON, text width=3.9cm, yshift=0.05cm, xshift=-0cm] {$S_{max_j}=\alpha \times \Std({P_{im}}_j(t_k))$};
\node (ESS_param) [process, label={[xshift=-2.9cm, yshift=-0cm]\textbf{Step 4}}, below of = find_Smax, text width=6.3cm, yshift = -0.05cm] {Set $C_{max_j}=\nicefrac{S_{max_j}}{\eta_{c_j}}$ and $D_{max_j}=\eta_{d_j} S_{max_j}$};
\node (policy) [process, label={[xshift=-2.85cm, yshift=-0.05cm]\textbf{Step 5}}, below of = ESS_param, text width=6.2cm, yshift = -0.05cm] {Apply optimal policy \eqref{policy1}--\eqref{policy2} to smooth ${P_{im}}_j(t_k)$};
\node (xi) [process, below of = policy, label={[xshift=-2.15cm, yshift=-0.04cm]\textbf{Step 6}}, yshift=-0.06cm, fill = cyan!10] {Obtain $\bm{x}=[\Delta\bm{\delta},\Delta\bm{\omega}]^T$ from PMU data};
\node (state_matrix) [process, label={[xshift=-1.65cm, yshift=-0.06cm]\textbf{Step 7}}, below of=xi, text width=3.8cm, yshift=-0.0cm, fill = cyan!10] {Estimate state matrix $A$ by \eqref{eq:estA_c}};
\node (IR_modes) [process, label={[xshift=-2.5cm, yshift=-0cm]\textbf{Step 8}}, below of=state_matrix, text width=5.5cm, yshift=-0.15cm, fill = cyan!10] {Perform modal analysis on $A$ and obtain the inter-area mode properties $f_i, \zeta_i, i\in \{1,...,n\}$};
\draw [arrow] (wind_speed) -- (check_stand);
\draw [arrow] (check_stand) -- node[anchor=west, xshift=0cm, yshift=0.05cm] {Yes} (ESS_ON);
\draw [arrow] (ESS_ON) -- (find_Smax);
\draw [arrow] (find_Smax) -- (ESS_param);
\draw [arrow] (ESS_param) -- (policy);
\draw [arrow] (policy) -- (xi);
\draw [arrow] (xi) -- (state_matrix);
\draw [arrow] (state_matrix) -- (IR_modes);
\draw [arrow] (check_stand) -- node[anchor=south, xshift=-0.1cm, yshift=0cm] {No} (ESS_OFF);
\draw [arrow] (ESS_OFF) |- (xi);
\end{tikzpicture}
\caption{Flowchart of the proposed ESS algorithm.}
\label{chart}
\end{center}
\vspace{-18pt}
\end{figure}
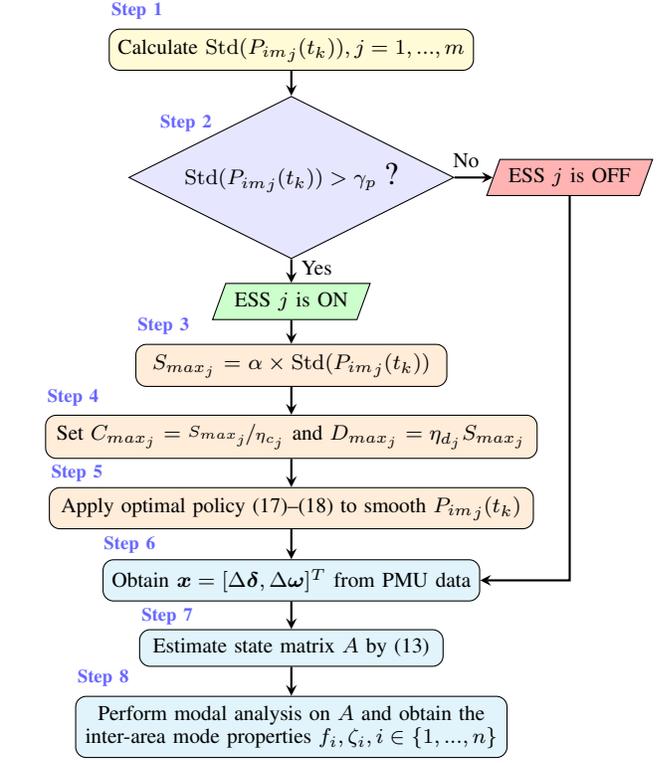

\noindent $\bullet$    In \textbf{Step 1},
the standard deviation
$\Std({P_{im}}_j(t_k))$  
    can be either calculated using the CDF of ${v_{w_j}}$ and \eqref{Pwind} or directly obtained from wind power data or a probabilistic model \cite{Gamal13}.\\
    \noindent $\bullet$   In \textbf{Step 2}, $\gamma_p$ is selected as a threshold for the acceptable deviations in the varying wind farm power output. 
     In this paper, $\gamma_p=0.1$ is chosen, 
     the suitability of which is confirmed by the numerical experiments (see Section \ref{4}). \\
    \noindent $\bullet$  In \textbf{Step 3}, the value of $\alpha$ is determined based on the relevant ESS literature \cite{Gamal13} and our simulation experience. 
    To reduce the expected average magnitude of the initial power imbalance, i.e. $\mu_{P_{im_j}}=\E(\frac{1}{N}\sum_{k=1}^{N}|P_{im_j}(t_k)|)$,
    by $70\%$, $S_{max_j}$ is set to be only $7\times \Std({P_{im}}_j(t_k))$, i.e. $\alpha=7$. The resulting $S_{max_j}$ is only 1 per unit (p.u.).\\ 
  \noindent $\bullet$  In \textbf{Step 4}, the efficiencies ${\eta_{c_j}}^2$ and ${\eta_{d_j}}^2$ typically lie in $[50\%,95\%]$ \cite{Gamal13} and are considered as pre-known parameters.
\section{Numerical Results}\label{4}

\begin{figure}[!t]
\centering
\includegraphics[width=2in ,keepaspectratio=true,angle=0]{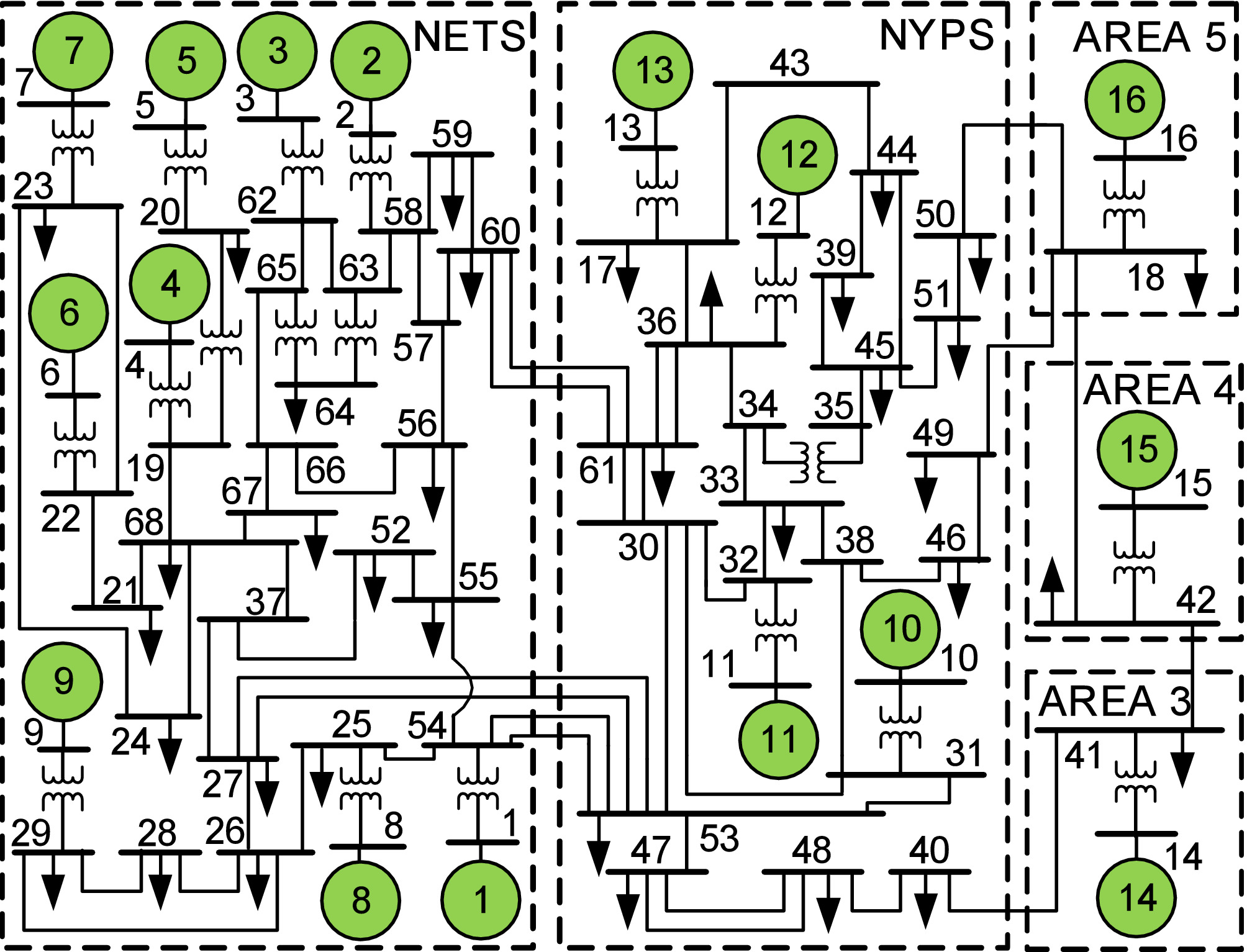}
\caption{68-bus, 16-generator, 5-area benchmark system \cite{Pal05}.}\label{NETS-NYPS_1}
\vspace{-11pt}
\end{figure}

In this section, the effectiveness of the proposed data-driven ESS strategy in enhancing
online
small-signal stability monitoring is validated. 
Simulations are conducted on the IEEE $68$-bus system; see Fig. \ref{NETS-NYPS_1}.
Detailed modal analysis reported in the literature \cite{Zenelis20} reveals the presence of three inter-area modes with typical frequencies $f_1=0.42$ Hz (mode $1$), $f_2=0.63$ Hz (mode $2$) and $f_3=0.77$ Hz (mode $3$).
The classical model has been used to represent synchronous machines.
%
Loads are modeled by constant impedances experiencing Gaussian variations with $\sigma_i=20$ \cite{Zenelis20}.
The rest of the power system is represented according to \cite{Pal05}. 
Wind power is integrated into the grid through the widely used doubly-fed induction generator (DFIG)
\cite{Slootweg03}.
The stochasticity of wind speed $\bm{v_w}$ is modeled by the Weibull distribution with shape parameter $k_{v_w}=1$ and scale parameter $\lambda_{v_w}=0.02$ obtained from real-life applications \cite{Minano13}.
Wind power fluctuations are smoothed using ESS with ${\eta_{c_j}}^2={\eta_{d_j}}^2=70\%$ and $S_{max_j}=\eta_{c_j} C_{max_j}=D_{max_j}/\eta_{d_j}=100$ MW ($1$ p.u.), $j=1,...,m$. Time-domain simulations are implemented in PSAT toolbox \cite{Milano05},
while 
ESS operates every $\nicefrac{1}{3}$ s \cite{Gamal13}.

\subsection{Validation of the Data-Driven ESS Algorithm for Enhancing Inter-Area Mode Identification} 
\label{4a}
In this case study, we install $500$ MW DFIG-based wind farms to the zero-injection buses $\{19,31,32,62\}$, reaching a wind penetration level of $11\%$. Buses $19,31,32,62$ correspond to $j=1,2,3,4$, respectively (see \eqref{eq_Pimb}). 
Next, we conduct $100$ Monte Carlo time-domain simulations (i.e. $100$ different wind speed realizations $\bm{v_w}$) to perform the probabilistic small-signal stability monitoring of the grid. Note that the proposed algorithm can work using a single scenario, yet Monte Carlo simulation results are more rigorous from a
statistical
sense.

Since $\Std(P_{im_j}(t_k))>\gamma_p=0.1$    p.u., $\forall j$, ESS is activated (ON). 
Table \ref{reduction} presents a comparison of the expected average magnitude of the power imbalance when the ESS is OFF (i.e.   $\mu_{P_{im_j}}$) and ON (i.e. $\mu_{P_{res_j}}$).
 It can be observed that the ESS algorithm
achieves a decrease of almost $70\%$ in the expected average magnitude of the power imbalance even though $S_{max_j}, j=1,...,4$ is very small given the system scale. This result is consistent with the findings of relevant works \cite{Gamal13}. Furthermore, the per-unit value of $\mu_{P_{res_j}}$ ($\approx 0.05$ p.u. $=5$ MW) approaches zero and is less than $1\%$ of the reference wind farm power output ($500$ MW $=5$ p.u.). 

 \begin{table}[!t]
\centering
  \caption{Wind Power Imbalance With and Without ESS}\label{reduction}
  \setlength{\tabcolsep}{5pt}
  \begin{tabular}{c c c c c}
\hhline{=====}
\hline
Bus&$j$&$\mu_{P_{im_j}}$ (p.u.) &$\mu_{P_{res_j}}$ (p.u.)&Decrease (\%)\Tstrut\Bstrut\\
  \hline
  $19$&$1$&$0.142$&$0.044$&$69.014$\Tstrut\\
  $31$&$2$&$0.138$&$0.046$&$66.667$\\
  $32$&$3$&$0.121$&$0.040$&$66.942$\\
  $62$&$4$&$0.132$&$0.041$&$68.939$\Bstrut\\
\hhline{=====}
  \end{tabular}
 \vspace{-12pt}
 \end{table}

\begin{table}[!t]
\centering
  \caption{Mean True and Estimated Properties of Inter-Area Mode $3$}\label{validation_mode_3}
  \setlength{\tabcolsep}{1.9pt}
  \begin{tabular}{c |c c c |c c c}
\hhline{=======}
\hline
ESS&$\E(f_{3_{t}})$ (Hz)&$\E(f_{3_{e}})$ (Hz)&Err. (\%)&$\E(\zeta_{3_{t}})$ (\%)&$\E(\zeta_{3_{e}})$ (\%)&Err.(\%)\Tstrut\Bstrut\\
  \hline
   OFF&$0.753$&$0.760$&$0.930$&$1.756$&$1.378$&$21.526$\Tstrut\\
   ON&$0.754$&$0.758$&$0.531$&$1.754$&$1.660$&$5.359$\Bstrut\\\hline
\hhline{=======}
  \end{tabular}\label{dampingperformance-1}

\raggedright{\scriptsize{Note: \qq{$_t$} stands for \qq{True}, \qq{$_{e}$} for \qq{Estimated}, and \qq{Err} for \qq{Error}. }}
\vspace{-12pt}
 \end{table}

Next, we compute the statistical properties of $A$, $f_i$ and $\zeta_i$, $i=1,2,3$.
To this end, we use $200$ s PMU data with a sampling frequency of $60$ Hz.
Simulation results show that wind stochasticity mostly affects the estimation of mode $3$. Therefore, a comparison of the mean true and estimated frequency and damping ratio of inter-area mode $3$ with and without ESS is presented in Table \ref{validation_mode_3}. Notice that, $\E(f_3)$ is accurately estimated when ESS is OFF despite wind uncertainty. On the other hand, the estimation of $\E(\zeta_3)$ without ESS is poor, thus deteriorating the small-signal stability monitoring.
These results demonstrate that the proposed data-driven ESS algorithm can reduce significantly the modal estimation error 
while requiring only a very small ESS capacity. 
Similar conclusions can be drawn for the other inter-area modes.


\subsection
{The 
Algorithm Operation under High Wind Penetration}

\begin{figure}[!t]
\centering
\includegraphics[width=2in ,keepaspectratio=true,angle=0]{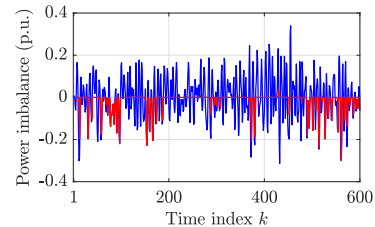}
\caption{Initial power imbalance ${P_{im_4}}(t_k)$ (displayed in blue) and residual power imbalance ${P_{res_4}}(t_k)$ (displayed in red) for the wind farm of bus $62$.}\label{Imbalance_fig}
\vspace{-15pt}
\end{figure}


In this case study, the effectiveness of the proposed ESS algorithm
is validated under higher wind power penetration levels. Particularly, we extend the base-case scenario of section \ref{4a} by installing additional DFIG-based wind farms of $500$ MW on top of the existing ones in buses $\{19,31,32,62\}$. Thus, three different wind penetration levels are considered:

\begin{enumerate}
    \item[(1)] Buses $\{19,32,31,62\}$ -- $11\%$ penetration (base case)
    \item[(2)] Buses $\{19,32,31,62,22,58\}$ -- $17\%$ penetration
    \item[(3)] Buses $\{19,32,31,62,22,58,35,43\}$ -- $23\%$ penetration
\end{enumerate}
while $\{22,58,35,43\}\rightarrow j =\{5,6,7,8\}$, respectively.
To visualise the effectiveness of the proposed ESS algorithm, Fig. \ref{Imbalance_fig} displays a comparison between the initial power imbalance (ESS is OFF) and the residual power imbalance (ESS is ON) of wind farm at bus $62$ ($j=4$), for a single realization of $\bm{v_w}$. With the exception of a few remaining power imbalance spikes, ${P_{res_4}}(t_k)$ becomes zero, thus resulting in an approximately constant wind farm power output.
Importantly, $S_{max_j}=100$ MW is sufficient to almost eliminate wind power variations despite the penetration increase.

Next, the accuracy of small-signal stability monitoring is assessed.
Again, modal analysis is carried out based on $100$ Monte Carlo simulations.  Table \ref{penetration_high} summarizes the comparison between the mean absolute percentage error (MAPE) in the estimation of the inter-area mode frequencies ($MAPE_f$) and damping ratios ($MAPE_\zeta$) when ESS is OFF and the corresponding MAPE when ESS is ON. For illustration purposes, the average of all three modes is computed.
As can be seen, frequency errors are trivial irrespective of the ESS application. Nonetheless, damping ratios, which are the main focus of small-signal stability monitoring, exhibit large estimation errors when ESS is deactivated, especially for higher wind penetration levels (e.g. $23\%$).
As a result, a poorly-damped inter-area mode can be potentially identified as well-damped. Clearly, our method promotes the accurate estimation of $\zeta_i$ by smoothing the unwanted wind power fluctuations.

\begin{table}[!t]
\centering
\caption{Mean Absolute Percentage Inter-Area Mode Estimation Error}\label{penetration_high}
\begin{tabular}{c c c c}
\hhline{====}
\hline
Wind Penetration&ESS&$MAPE_f$ (\%) & $MAPE_\zeta$ (\%)\Tstrut\Bstrut\\
  \hline
  $11\%$&OFF&$2.316$&$9.609$\Tstrut\\
        &ON&$2.914$&$2.653$\Bstrut\\\hline
  $17\%$&OFF&$1.653$&$9.652$\Tstrut\\
        &ON&$2.214$&$1.659$\Bstrut\\\hline
  $23\%$&OFF&$1.043$&$16.289$\Tstrut\\
        &ON&$1.260$&$3.337$\Bstrut\\
\hhline{====}
  \end{tabular}
 \vspace{-15pt}
 \end{table}

\section{Conclusion and Perspectives}\label{5}

This paper proposed a novel data-driven ESS algorithm for small-signal stability monitoring of power systems with stochastic wind power penetration. Our method can accurately estimate the inter-area mode properties by smoothing the wind power variations, thus enhancing the small-signal stability assessment in near real-time. Numerical simulations demonstrate that the proposed technique achieves the inter-area mode identification and smoothing goals using only a small ESS capacity. Future endeavors will focus on developing methodologies for optimal ESS placement in wind farms and on implementing the proposed data-driven ESS in practice. 
\vspace{-5pt}


\begin{thebibliography}{99}







































































\bibitem{Bu15}
S. Bu, W. Du, and H. Wang, \qq{Investigation on probabilistic small-signal stability of power systems as affected by offshore wind generation,} \textit{IEEE Trans. Power Syst.,} vol. 30, no. 5, pp. 2479-2486, Sep. 2015.

\bibitem{Bu12}
S. Bu, W. Du, H. Wang, Z. Chen, L. Xiao, and H. Li, \qq{Probabilistic analysis of small-signal stability of large-scale power systems as affected by penetration of wind generation,} \textit{IEEE Trans. Power Syst.,} vol. 27, no. 2, pp. 762-770, May 2012.

\bibitem{Andersson05}
G. Andersson \textit{et al.,} \qq{Causes of the 2003 major grid blackouts in North America and Europe,
and recommended means to improve system dynamic performance,} \textit{IEEE Trans.
Power Syst.,} vol. 20, no. 4, pp. 1922-1928, Nov. 2005.

\bibitem{Huang13}
H. Huang, C. Chung, K. Chan, and H. Chen, \qq{Quasi-Monte Carlo based probabilistic small signal stability analysis for power systems with plug-in electric vehicle and wind power integration,} \textit{IEEE Trans. Power Syst.,} vol. 28, no. 3, pp. 3335-3343, Aug. 2013.

\bibitem{Bazargan10}
C. Wang, L. Shi, L. Yao, L. Wang, Y. Ni, and M. Bazargan, \qq{Modelling analysis in power system small signal stability considering uncertainty of wind generation,} in \textit{Proc. IEEE PES General Meeting,} Providence, RI, 2010.

\bibitem{Sumper12}
F. Daz-Gonz\'{a}lez, A. Sumper, O. Gomis-Bellmunt, and R. Villaf\'{a}fila- Robles, \qq{A review of energy storage technologies for wind power applications,} \textit{Renewable Sust. Energy Rev.,} vol. 16, no. 4, pp. 2154–2171, May 2012.

\bibitem{Sheng20}
H. Sheng and X. Wang, \qq{Online measurement-based estimation of dynamic system state matrix in ambient conditions,} \textit{IEEE Trans. Smart Grid,} vol. 11, no. 1, pp. 95-105, Jan. 2020.


\bibitem{Zenelis18}
I. Zenelis and X. Wang, \qq{Wide-area damping control for interarea oscillations in power grids based on PMU measurements,} {\em IEEE Control Syst. Lett.,} vol. 2, no. 4, pp. 719-724, Oct. 2018.

\bibitem{Zenelis20}
I. Zenelis, X. Wang, and I. Kamwa, \qq{Online PMU-based wide-area damping control for
multiple inter-area modes}, \textit{IEEE Trans. Smart Grid,} vol. 11, no. 6, pp. 5451-5461, Nov. 2020.

\bibitem{Weng20}
Y. Hashmy, Z. Yu, D. Shi, and Y. Weng, \qq{Wide-area measurement system-based low frequency oscillation damping control through reinforcement learning,} \textit{IEEE Trans. Smart Grid,} vol. 11, no. 6, pp. 5072-5083, Nov. 2020.

\bibitem{Kamwa01}
I. Kamwa, R. Grondin, and Y. Hebert, \qq{Wide-area measurement-based stabilizing control of large power systems--A decentralized/hierarchical approach,} {\em IEEE Trans. Power Syst.}, vol. 16, no. 1, pp. 136-153, Feb. 2001.









\bibitem{PNNL}
\textit{Wide-Area energy storage and management system to balance intermittent resources in the Bonneville Power Administration and California ISO control areas}, Pacific Northwest National Laboratory, Tech. Rep., Jun. 2008, [Online]. Available: https://www.osti.gov/biblio/947483

\bibitem{Hasanien14}
H. Hasanien, \qq{A set-membership affine projection algorithm-based adaptive-controlled SMES units for wind farms output power smoothing,} \textit{IEEE Trans. Sustain. Energy,} vol. 5, no. 4, pp. 1226-1233, Oct. 2014.



\bibitem{Kundur94}
P. Kundur, \textit{Power System Stability and Control.} New York, NY, USA: McGraw-Hill, 1994.





\bibitem{Theresa13}
T. Odun-Ayo and M. Crow, \qq{An analysis of power system transient stability using stochastic energy functions,} {\em Int. Trans. Electr. Energ. Syst.,} vol. 23, no. 2, pp. 151-165, Oct. 2011.

\bibitem{Carta09}
J. Carta, P. Ram\'{i}rez, and S. Vel\'{a}zquez, \qq{A review of wind speed probability distributions used in wind energy analysis: Case studies in the Canary Islands,}
\textit{Renewable Sust. Energy Rev.,} vol. 13, no. 5, pp. 933-955, Jun. 2009.

\bibitem{Milano13}
F. Milano and R. Z\'{a}rate-Mi\~{n}ano, \qq{A systematic method to model power systems as stochastic differential algebraic equations,} \textit{IEEE Trans. Power Syst.}, vol. 28, no. 4, pp. 4537-4544, Nov. 2013.

\bibitem{Wang15}
X. Wang, H. Chiang, J. Wang, H. Liu, and T. Wang, \qq{Long-term stability analysis of power systems with wind power based on stochastic differential equations: model development and foundations,} \textit{IEEE Trans. Sustain. Energy,} vol. 6, no. 4, pp. 1534-1542, Oct. 2015.

\bibitem{Pierrou19}
G. Pierrou and X. Wang, \qq{The effect of the uncertainty of load and renewable generation on the dynamic voltage stability margin,} in \textit{Proc. IEEE PES Innov. Smart Grid Technol. Eur.}, Bucharest, Romania, 2019.

\bibitem{Slootweg03}
J. Slootweg, \textit{Wind power: modelling and impact on power system dynamics,}, Delft University of Technology, 2003.

\bibitem{Zhou15}
N. Zhou, D. Meng, Z. Huang, and G. Welch, \qq{Dynamic state estimation of a synchronous machine using pmu data: A comparative study,} {\em IEEE
Trans. Smart Grid,} vol. 6, no. 1, pp. 450-460, Jan. 2015.



































































\bibitem{Gamal13}
H. Su and A. Gamal, \qq{Modeling and analysis of the role of energy storage for renewable integration: power balancing,} \textit{IEEE Trans. Power Syst.,} vol. 28, no. 4, pp. 4109-4117, Nov. 2013.

\bibitem{Hug10}
G. Hug, \qq{Coordination of intermittent generation with storage, demand control and conventional energy sources}. in \textit{Proc. IREP Symposium-Bulk Power System Dynamics
and Control-VIII,} Buzios, Brazil, 2010.

\bibitem{Pal05}
B. Pal and B. Chaudhuri, \textit{Robust Control in Power Systems}. New York, NY, USA: Springer, 2005.


\bibitem{Minano13}
R. Z\'{a}rate-Mi\~{n}ano, M. Anghel, and F. Milano, \qq{Continuous wind speed models based on stochastic differential equations,} \textit{Appl. Energy,} vol. 104, no.1, pp. 42–49, Apr. 2013.






































\bibitem{Milano05}
F. Milano, \qq{An open source power system analysis toolbox,} \textit{IEEE Trans.
Power Syst.,} vol. 20, no. 3, pp. 1199-1206, Aug. 2005.






\end{thebibliography}
\end{document}